\documentclass[aps,prc,twocolumn,groupedaddress,showpacs]{revtex4}
\usepackage{epsfig}
\usepackage{dcolumn}% Align table columns on decimal point

\begin{document}

%\preprint{}

\title{Isoscaling in Light-Ion Induced Reactions
and its Statistical Interpretation}

\author{A.S.~Botvina,$^{(1)}$\cite{AAA}
O.V.~Lozhkin,$^{(2)}$
and W.~Trautmann$^{(1)}$
}

\affiliation{$^{(1)}$Gesellschaft  f\"ur  Schwerionenforschung, D-64291 
Darmstadt,
Germany\\
$^{(2)}$V.G. Khlopin Radium Institute, 197002 St. Petersburg, Russia
}

%\date{\today}

\begin{abstract}

Isotopic effects observed in fragmentation reactions induced by
protons, deuterons, and $\alpha$ particles of incident energies
between 660~MeV and 15.3~GeV on $^{112}$Sn and $^{124}$Sn targets
are discussed. The exponential scaling of the yield ratios with the
third component of the fragment isospin $t_3 = (N-Z)/2$ is observed
in all reactions, with scaling parameters that depend on the incident
energy.
Breakup temperatures for these reactions are deduced from double ratios
of isotopic yields and tested for their relation with the isoscaling
parameters.

The quantum statistical (QSM) and the statistical multifragmentation (SMM)
models are used for interpreting the results.
The observed isoscaling can be understood as a consequence of a statistical
origin of the emitted fragments in these reactions.
The SMM analysis shows that the exponent describing the isoscaling
behavior is proportional to the strength of the
symmetry term of the fragment binding energy.
Using this result, a symmetry-term coefficient $\gamma \approx$ 22.5 MeV
for fragments at breakup is deduced from the experimental data.
This is close to the standard value and supports SMM assumptions for the
breakup configuration.
An alternative method of
determining the symmetry-energy coefficient, by using isotope distribution
widths, is also discussed.
\end{abstract}

%\keywords{}

\maketitle

\section{\label{sec:intro}Introduction}

Isotopic effects in nuclear reactions are receiving increasing attention
because of their relation with the symmetry energy in the nuclear
equation of state whose density dependence is of high current interest, 
in particular also for astrophysical applications 
\cite{muell95,bali98,tan01,bali01}.
In a series of recent papers, the scaling properties of cross sections for 
fragment production with respect to the isotopic composition of the 
emitting systems were investigated by Tsang et al. 
\cite{tsang01,tsang01a,tsang01b}. The studied reactions include
symmetric heavy-ion reactions at intermediate energy leading to 
multifragment-emissions as well as asymmetric reactions induced by $\alpha$
particles and $^{16}$O projectiles at low to intermediate energies with
fragment emission from excited heavy residues.
The common behavior observed for these reactions, termed isoscaling,
concerns the production ratios $R_{21}$ for fragments with neutron number 
$N$ and proton number $Z$ in reactions with different isospin asymmetry.
It is constituted by 
their exponential dependence on $N$ and $Z$ according to
\begin{equation}
R_{21}(N,Z) = Y_2(N,Z)/Y_1(N,Z) = C \cdot exp(N\cdot \alpha + Z\cdot
\beta)
\label{eq:scalab}
\end{equation}
with three parameters C, $\alpha$ and $\beta$.
Here $Y_2$ and $Y_1$ denote the yields from the more neutron rich 
and the more neutron poor reaction system, respectively.

In some of the reactions, the parameters $\alpha$ and $\beta$ 
have the tendency to be
quite similar in absolute magnitude but of opposite sign. For 
multifragmentation following central collisions 
of $^{124}$Sn + $^{124}$Sn and $^{112}$Sn + $^{112}$Sn at 50 MeV 
per nucleon, $\alpha$ = 0.37 and $\beta$ = -0.40 was obtained from fits
to the fragment yield ratios in the mass range $1 \le A \le 18$ 
\cite{tsang01}. These parameters suggest an approximate
scaling with the third component of the isospin $t_3 = (N-Z)/2$
of the form 
\begin{eqnarray}
R_{21}(N,Z) = C \cdot exp((N-Z)\cdot \alpha) \cdot exp(Z\cdot(\alpha+
\beta)) \nonumber\\
\approx C \cdot exp(t_3 \cdot 2\alpha),
\label{eq:scalt3}
\end{eqnarray}
which follows from Eq.~(\ref{eq:scalab}) since $\beta \approx -\alpha$.

The isotopic scaling of this latter
kind has first been reported for reactions of 
protons of 660~MeV energy incident on targets of $^{112,124}$Sn 
\cite{boga74,boga76} and subsequently also for other reactions with 
proton and deuteron projectiles
in the relativistic regime of bombarding energies \cite{boga80}. 
With the notation chosen in these early papers,
\begin{equation}
R_{12}(N,Z) = Y_1(N,Z)/Y_2(N,Z) = C \cdot exp(-t_3 \cdot \beta_{t3}),
\label{eq:scalboga}
\end{equation}
parameter values in the range $\beta_{t3} \approx 0.7$ were obtained for the 
reactions with p(6.7 GeV) and d(3.1 GeV) projectiles \cite{notation}. 
This is rather
close to twice the value of $\alpha$ for the symmetric Sn + Sn reactions
\cite{tsang01} and, according to Eq.~(\ref{eq:scalt3}), suggests 
a very similar scaling behavior for fragmentation reactions
induced by relativistic light ions.

In this paper, we will discuss the isotopic effects observed by 
Bogatin et al. for reactions induced by 
protons, deuterons, and $\alpha$ particles of incident energies 
between 660~MeV and 15.3~GeV on $^{112,124}$Sn targets
\cite{boga74,boga76,boga80,boga82}. Particular emphasis will be given to
their scaling properties, with the aim to incorporate the
light-ion induced fragmentation 
into the set of reactions investigated by Tsang et al. \cite{tsang01}.
A complete set of references to these data and a statistical analysis 
performed with the quantum statistical model (QSM) of Hahn and St\"ocker
\cite{hahn88} can be found in Ref.~\cite{lozh92}.

We will, furthermore, present isotopic temperatures derived from double
ratios of helium and lithium isotopes for these reactions 
and compare their dependence on the
incident energy with that of the scaling parameters.
Temperature measurements, in principle, also permit a test whether
the reaction scenario, and specifically the temperature as an important 
parameter characterizing it, are indeed independent of the isotopic 
composition of the system as commonly assumed. For isotope temperatures,
however, this property is already implied if isoscaling holds.
It is a consequence 
of Eq.~(\ref{eq:scalab}) according to which the double yield ratios from which 
isotope temperatures are derived are identical for the pair of reactions.
The observations of approximately identical isotope temperatures 
in the $^{112}$Sn + $^{112}$Sn
and $^{124}$Sn + $^{124}$Sn reactions at 50 MeV per nucleon \cite{kunde98},
as well as for the present reactions, are therefore part of the 
more general phenomenon of isoscaling.

It has been shown that in both, light-ion induced collisions and 
peripheral heavy-ion collisions at high energy, the fragment production and 
observed isotopic effects can be explained in the framework of a 
hybrid approach consisting of a dynamical 
initial stage and a subsequent statistical breakup 
of a highly excited residual at low density
\cite{bond95,botv95,avde98,wang99}.  
With the aim to identify reasons for the isotopic scaling in the
present case, an analysis with the statistical multifragmentation 
model (SMM, Ref. \cite{bond95}) was carried out. It will be demonstrated
that isotopic scaling arises naturally in a statistical fragmentation 
mechanism.
The isoscaling parameter $\alpha$ deduced for hot primary fragments 
is, furthermore, found to be directly proportional to the symmetry 
part of the binding energy of the fragments when they are formed 
at low density. To the extent that the modification of this parameter 
during secondary deexcitation remains small this opens the possibility 
of testing
components of the nuclear equation of state in fragmentation reactions.

\section{\label{sec:exper}Experimental data}

The experimental data are taken from the literature 
\cite{boga74,boga76,boga80,boga82}. They were obtained in the JINR 
laboratories in Dubna with beams of protons of 660~MeV, 1.0 and 6.7~GeV,
of deuterons with 3.1~GeV, and of $\alpha$ particles with 15.3~GeV incident 
energy. Isotopically resolved cross sections of light fragments were 
measured with semiconductor telescopes placed at 
$\theta_{lab} = 90^{\circ}$ and with thin internal targets 
made from enriched $^{112}$Sn ($\approx$ 81\%) and $^{124}$Sn 
($\approx$ 96\%). From the yields, 
integrated over energy intervals specified in Refs.~\cite{boga74,boga80}, 
ratios $R_{12}$ for the production of a particular fragment in the reactions
with the two Sn isotopes were determined.
It is convenient to introduce
a reduced isotopic effect for a fragment species X by normalizing 
with respect to the ratio observed for $^6$Li, i.e. 
$R_{12}(X)/R_{12}(^6{\rm Li})$. Uncertainties of the 
absolute normalizations of the data sets measured with the two targets 
are thus eliminated.

\begin{figure}
     \epsfysize=9.0cm
     \centerline{\epsffile{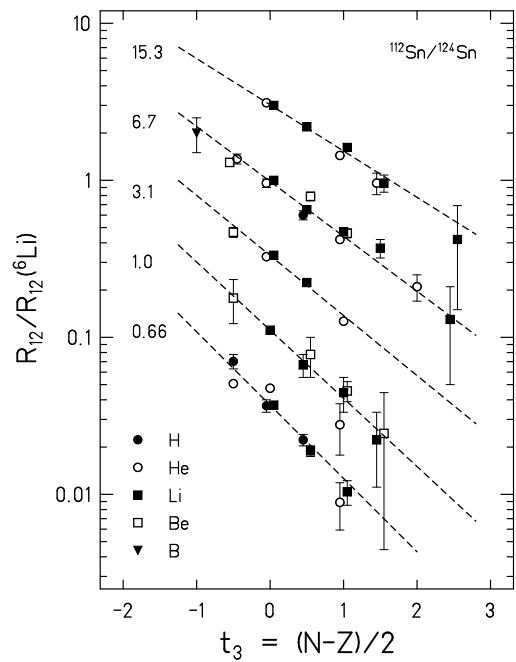}}
\caption{\label{fig:fig1} 
Isotopic effect $R_{12}$, normalized with respect to $R_{12}(^6{\rm Li}$),
versus the third component of the fragment isospin $t_3$. The five
reactions are offset from each other by multiple factors of three and
are labelled with the total projectile energy, given in units of
GeV. H, He, Li, Be, and B fragments are distinguished by
different data symbols as indicated. The lines are the results of
exponential fits according to Eq.~(\protect\ref{eq:scalboga}).
Some of the data symbols are slightly displaced
horizontally for reasons of clarity.
}
\end{figure}

The reduced isotopic effects measured for the five pairs of 
reactions are shown in Fig.~\ref{fig:fig1}. The
cross section ratios for the most neutron poor and the most neutron
rich fragments differ by about one order of magnitude 
in all cases except for the d(3.1 GeV) reaction
for which only a narrow range of isospin is covered by the detected
products. A nearly perfect exponential dependence
on the third component $t_3$ of the fragment isospin is observed,
with slope parameters $\beta_{t3}$ (Eq.~(\ref{eq:scalboga})) 
that decrease 
gradually from 1.08 to 0.68 as the projectile energy increases 
(Table~\ref{tab:table1}).
This variation of the isotopic effect with the incident energy has been
noted in Ref.~\cite{lozh92} and tentatively ascribed to a gradual rise of the
temperatures of the emitting systems. 

\begin{table}
\caption{\label{tab:table1}
Parameters obtained from fitting the measured isotopic
yield ratios with the scaling functions given in
Eqs.~(\protect\ref{eq:scalab}) and (\protect\ref{eq:scalboga}).
The second column gives the range of fragment $Z$ over which the
data sets extend.
}
\begin{ruledtabular}
\begin{tabular}{l c c c c}
%\hline\hline
Projectile & $Z$ & $\beta_{t3}$ & $\alpha$ & $\beta$ \\
\hline
p 0.66 GeV &
 1 - 3 &
 1.08 $\pm$ 0.06 &
 0.53 $\pm$ 0.04 &
-0.51 $\pm$ 0.05 \\
%\hline
p 1.00 GeV &
 2 - 4 &
 1.00 $\pm$ 0.10 &
 0.52 $\pm$ 0.04 &
-0.65 $\pm$ 0.05 \\
%\hline
d 3.10 GeV &
 2 - 4 &
 0.88 $\pm$ 0.04 &
 0.43 $\pm$ 0.03 &
-0.45 $\pm$ 0.04 \\
%\hline
p 6.70 GeV &
 1 - 5 &
 0.81 $\pm$ 0.02 &
 0.39 $\pm$ 0.01 &
-0.43 $\pm$ 0.02 \\
%\hline
$\alpha$ 15.3 GeV &
 2 - 3 &
 0.68 $\pm$ 0.02 &
 0.34 $\pm$ 0.01 &
-0.32 $\pm$ 0.03 \\
%\hline\hline
\end{tabular}
\end{ruledtabular}
\end{table}

\begin{figure}
     \epsfysize=8.0cm
     \centerline{\epsffile{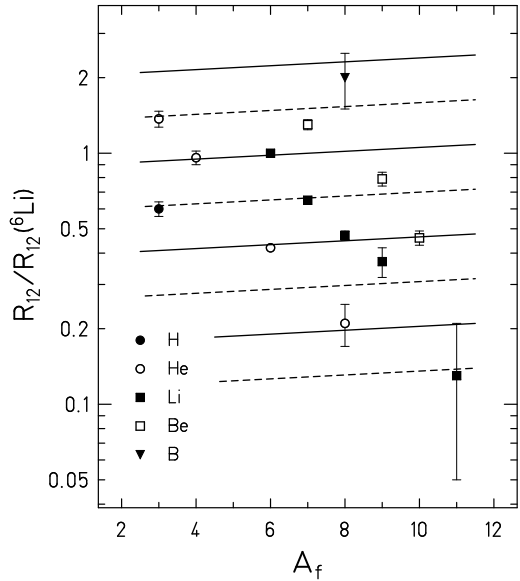}}
\caption{\label{fig:fig2} 
Isotopic effect $R_{12}$, normalized with respect to $R_{12}(^6{\rm Li}$),
versus the mass number $A_{\rm f}$ of the detected fragment for the
reactions of protons with $^{112,124}$Sn at 6.7 GeV.
H, He, Li, Be, and B fragments are distinguished by
different data symbols as indicated.
The result of a three-parameter fit to the data according
to Eq.~(\protect\ref{eq:scalat3}) is represented by the lines of
constant integer (full lines) and half-integer (dashed) isospin.
}
\end{figure}

Two-parameter fits according to Eq.~(\ref{eq:scalab}) were also performed,
with results that are listed in Table~\ref{tab:table1}. 
The monotonic trend exhibited by
the parameter $\alpha$ as a function of the incident energy
reflects that of $\beta_{t3}$. It apparently extends to much lower 
energies, as evident from the value $\alpha$ = 0.60 reported 
for the $\alpha$(200 MeV) reaction in Ref.~\cite{tsang01}.
The dependence on
$Z$ is not equally well established for all reactions since only a
limited range of elements has been covered in some cases. There is,
however, a tendency of the absolute value of $\beta$ being larger 
than $\alpha$. For protons of 6.7 GeV, this is illustrated 
in Fig.~\ref{fig:fig2} with the results from yet another 
parameterization in terms of $A$ and $t_3$,
\begin{equation}
R_{12}(A,t_3) = C \cdot exp(A\cdot \alpha_A + t_3\cdot \beta'_{t3}).
\label{eq:scalat3}
\end{equation}
The logarithmic slope of the cross section ratios 
for given $t_3$ as a function of $A$, by its definition 
equal to half the difference between $|\beta|$ and $\alpha$, is finite
with a value $\alpha_A = (1.8 \pm 1.1) \cdot 10^{-2}$.
Weighted over the five reactions $|\beta|$ is found to be
larger than $\alpha$ by 8\% $\pm$ 4\%.  

\section{\label{sec:dynam}Initial dynamical stage}

For the simulation of the initial stage of the collision
the intranuclear cascade 
(INC) model developed in Dubna was used \cite{toneev83,botv90}. 
The INC describes the process of the hadron-nucleon collisions inside 
the target nucleus. High energy products of these interactions are
allowed to escape 
while low energy products are assumed to be trapped by the nuclear
potential of the target system. At the end of the cascade, a residue
with a certain mass, charge and excitation energy remains which then
can be used as input for the statistical description of the fragment 
production.

\begin{figure}
     \epsfysize=8.0cm
     \centerline{\epsffile{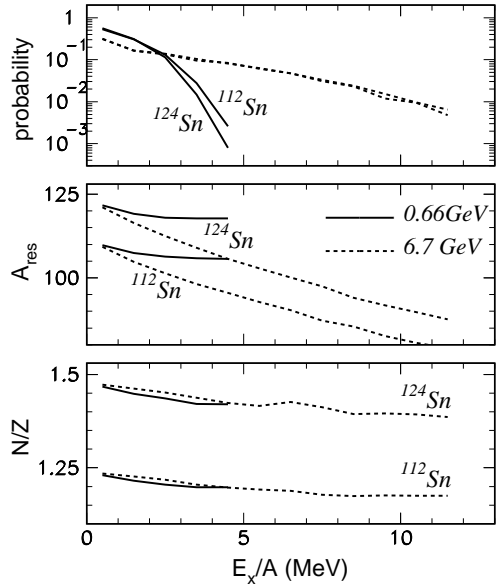}}
\caption{\label{fig:fig3} 
Production probability of residual nuclei after the intranuclear
cascade (top), their mean mass numbers $A_{\rm res}$ (middle), and their
mean neutron-to-proton ratio $N/Z$ (bottom) as a function of
their excitation energies for collisions of protons with
$^{112}$Sn and $^{124}$Sn targets at 660~MeV (solid lines) and
6.7~GeV (dashed lines).
}
\end{figure}

For p + $^{112,124}$Sn reactions,
the obtained correlations between the 
mass number and the $N/Z$ ratio of the residues with their
excitation energy is shown in Fig.~\ref{fig:fig3}. 
The masses decrease with increasing 
excitation energy, a behavior that is well known 
\cite{bond95,botv90,gaim91,poch95},
but the rate is considerably lower for the
lower proton energy. The $N/Z$ ratio also decreases gradually for both 
targets with an apparently universal rate that does not depend much on 
the projectile energy nor on the neutron content of the target. 
As a consequence, the difference $\Delta (N/Z)$ on which the isotopic 
effect depends linearly in first order \cite{tan01,boga76}
remains approximately constant. The calculated cross sections show that the 
covered range of excitation energies depends strongly on the proton energy.

It has been noticed repeatedly that the excitation energies obtained 
from first-stage reaction models are larger than needed to describe 
the observed fragment production with statistical multifragmentation models
\cite{avde98,botv92,barz93,bao93,schuet96}. 
This effect has been interpreted as 
evidence for expansion and additional preequilibrium emission during an 
intermediate stage between the cascade termination
and the fragment formation, not accounted for in the two-stage 
description. It leads to an uncertainty for the input parameters of 
the statistical calculations which has to be considered in their use 
and interpretation.

\section{\label{sec:chemi}Chemical equilibrium}

Isotopic effects and isotope yield ratios confront us with
the question of chemical equilibrium in the system.
Here, the grand-canonical 
quantum-statistical models (QSM) are useful for extracting 
relative isotopic abundances that correspond to the 
thermodynamical limit.
The model of Hahn and St\"ocker \cite{hahn88}, chosen in the present case,
assumes thermal and chemical equilibrium at the breakup point where the
fragmenting system is characterized by a density $\rho$, temperature $T$,
and by its overall $N/Z$ ratio. The model respects fermion and boson
statistics which, however, is not crucial at high temperature. It does not
take into account the finite size of the nuclear systems nor 
the Coulomb interaction between fragments but follows the
sequential decay of excited fragments according to tabulated
branching ratios. It has already been 
shown that the $t_3$ scaling (Eq.~(\ref{eq:scalboga})) exhibited by 
the p(6.7 GeV) + $^{112,124}$Sn reactions is well reproduced 
by the QSM if appropriate parameters are chosen \cite{lozh92}.
Even if $\Delta (N/Z)$ is fixed, e.g. with the aid of the INC model,
a continuous set of pairs of $T-\rho$ parameters can be found that
all permit equally good descriptions of the data. 
By varying either the temperature or the density the observed variation 
of the scaling parameter with incident energy can be followed.

\begin{figure}
     \epsfysize=7.0cm
     \centerline{\epsffile{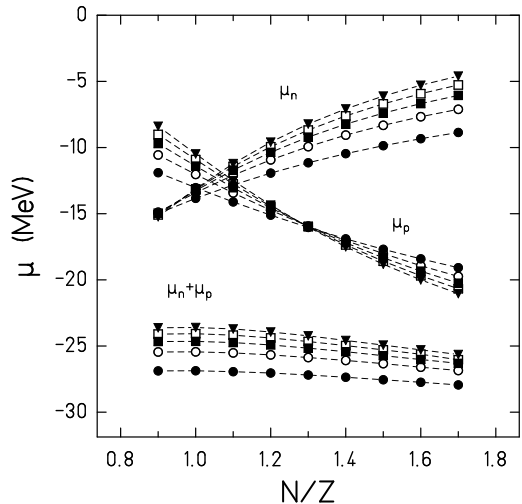}}
\caption{\label{fig:fig4} 
Results of QSM calculations for neutron and proton chemical potentials
$\mu_{\rm n}$ and $\mu_{\rm p}$ of
systems with different densities $\rho$ and $N/Z$ ratios.
The temperature is $T=6$~MeV;
the density increases from $\rho /\rho_0=0.1$ (dots) to 0.5 (triangles)
in steps of 0.1 where $\rho_0$ is the normal nuclear density.
}
\end{figure}

In the grand-canonical approximation,
the scaling parameters $\alpha$ and $\beta$ (Eq.~(\ref{eq:scalab})) are 
equal to the difference of the chemical potentials for neutrons and 
protons in the two systems, $\alpha = \Delta \mu_n /T$ and 
$\beta = \Delta \mu_p/T$, provided a common temperature $T$ for both 
systems exists \cite{tsang01,albergo}. The observation of $t_3$ scaling,
consequently, implies that these differences are of different sign and
about equal magnitude, or that the sum of $\mu_n$ and $\mu_p$ is invariant 
with the $N/Z$ ratio of the system. This is illustrated in 
Fig.~\ref{fig:fig4} in which
the chemical potentials extracted from the model calculations
are given as a function of $N/Z$. The chosen parameters are $T$ = 6 MeV
and $\rho /\rho_0$ from 0.1 to 0.5 in steps of 0.1 (the p(6.7 GeV) data,
e.g., are reproduced with $T$ = 6 MeV and $\rho /\rho_0$ = 0.1
\cite{lozh92}). The sum 
$\mu_n + \mu_p$ is approximately independent of $N/Z$,
with a very small tendency of
$\mu_p$ to change more rapidly than $\mu_n$.

For the Sn isotopes with $N/Z$ = 1.24 and 1.48 and for a breakup 
density $\rho /\rho_0$ = 0.3 the calculated differences of the chemical 
potentials are $\Delta \mu_n$ = 2.3~MeV and $\Delta \mu_p$ = -2.9~MeV. 
From these values coefficients $\alpha$ = 0.38 and $\beta$ = -0.48 
are obtained which are not far from the experimental observation 
in central Sn + Sn collisions \cite{tsang01} and in some of the present 
reactions. A more stringent model test will have to include
a comparison with fragment yields and an accurate estimation of the 
temperature. However, as also shown in Ref.~\cite{lozh92}, 
the chemical equilibrium hypothesis is quite adequate for the description 
of isotopic phenomena in these reactions
even though the heavy fragments or residues in the final channels 
are not explicitly taken into account. These degrees of freedom will be 
included in the SMM analysis presented in Section~\ref{sec:smmin}.

\section{\label{sec:tempe}Temperatures}

\begin{table}
\caption{\label{tab:table2}
Apparent temperatures deduced from He and Li isotopic
yield ratios, as indicated in column 2, for reactions with the
$^{112}$Sn and $^{124}$Sn targets.
}
\begin{ruledtabular}
\begin{tabular}{l c c c}
%\hline\hline
Projectile & isotopes & $^{112}$Sn & $^{124}$Sn \\
\hline
p 0.66 GeV & $^{3,4}$He, $^{6,7}$Li & 3.8 $\pm$ 0.1 MeV & 4.6 $\pm$ 0.1 MeV \\
\hline
p 1.00 GeV & $^{4,6}$He, $^{6,8}$Li & 2.2 $\pm$ 0.2 MeV & 2.6 $\pm$ 0.2 MeV \\
d 3.10 GeV &           "            & 2.7 $\pm$ 0.2 MeV & 3.1 $\pm$ 0.2 MeV \\
p 6.70 GeV &           "            & 2.9 $\pm$ 0.2 MeV & 2.9 $\pm$ 0.2 MeV \\
$\alpha$ 15.3 GeV &           "            & 3.3 $\pm$ 0.2 MeV 
& 3.5 $\pm$ 0.2 MeV \\
\hline
p 6.70 GeV & $^{6,8}$He, $^{6,8}$Li & 4.6 $\pm$ 0.4 MeV & 4.7 $\pm$ 0.3 MeV \\
$\alpha$ 15.3 GeV &           "            & 4.3 $\pm$ 0.5 MeV 
& 3.8 $\pm$ 0.3 MeV \\
%\hline\hline
\end{tabular}
\end{ruledtabular}
\end{table}

The reported cross sections for helium and lithium isotopes were used 
to construct temperature observables from double-isotope ratios
\cite{albergo} for the present set of reactions.
The production of $^{3,4}$He and of $^{6,7}$Li
has been measured for incident protons of 660~MeV, and the 
frequently used $T_{\rm HeLi}$ temperature \cite{poch95} can be 
determined for this particular case. Cross sections for the production 
of $^{3}$He are not reported for the reactions at higher
energies, so that $T_{\rm HeLi}$ cannot be used to follow the evolution
of the breakup temperature with incident energy. 

A common 
temperature observable for four out of the set of five reactions 
can be obtained from the $^{4}$He/$^{6}$He and $^{6}$Li/$^{8}$Li yield 
ratios, and in two cases $^{6}$He and $^{8}$He yields are available
which can also be combined with the lithium ratios $^{6}$Li/$^{8}$Li.
The corresponding expressions are
\begin{equation}
T_{\rm HeLi,0} = 13.3 MeV/\ln(2.2\frac{Y_{^{6}{\rm Li}}/Y_{^{7}{\rm Li}}}
{Y_{^{3}{\rm He}}/Y_{^{4}{\rm He}}});
\label{eq:theli}
\end{equation}

\begin{equation}
T_{{\rm He}46/{\rm Li}68,0} = -8.3 MeV/\ln(1.4\frac{Y_{^{6}{\rm Li}}/
Y_{^{8}{\rm Li}}}{Y_{^{4}{\rm He}}/Y_{^{6}{\rm He}}});
\label{eq:t4668}
\end{equation}

\begin{equation}
T_{{\rm He}68/{\rm Li}68,0} = 7.2 MeV/\ln(1.7\frac{Y_{^{6}{\rm Li}}/
Y_{^{8}{\rm Li}}}{Y_{^{6}{\rm He}}/Y_{^{8}{\rm He}}}).
\label{eq:t6868}
\end{equation}

\begin{figure}
     \epsfysize=8.0cm
     \centerline{\epsffile{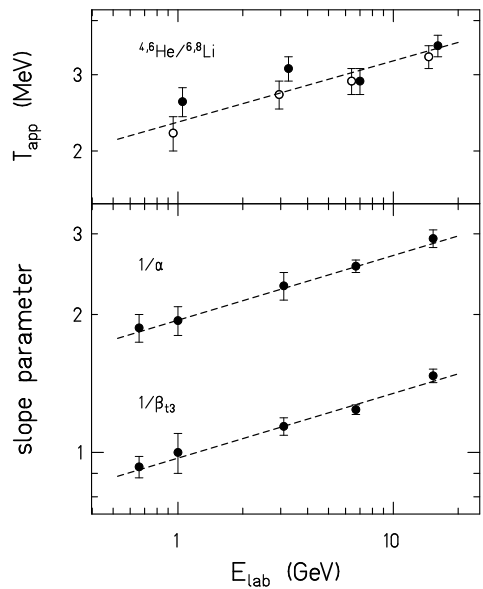}}
\caption{\label{fig:fig5} 
Apparent isotope temperature $T_{\rm app}$ deduced from $^{4,6}$He and
$^{6,8}$Li yield ratios (top) and the inverse isoscaling parameters
(bottom) 1/$\alpha$ (Eq.~(\protect\ref{eq:scalab})) and 1/$\beta_{t3}$
(Eq.~(\protect\ref{eq:scalboga})) as a function of the total projectile 
energy.
The dashed lines represent the logarithmic rise of
1/$\beta_{t3}$, multiplied by appropriate factors for
comparison with the trends observed for 1/$\alpha$ and $T_{\rm app}$.
}
\end{figure}

The two latter isotopic thermometers do not fulfill the requirement that 
the double difference of the binding energies of the four 
isotopes, the prefactor in Eqs.~(\ref{eq:theli}) - (\ref{eq:t6868}), should 
be large compared to the anticipated temperatures \cite{poch95,tsang97}.
They may thus be more strongly influenced by sequential decays.
In particular, the contributions from residue evaporation to
the inclusive yields of $^4$He will have a large effect on
$T_{{\rm He}46/{\rm Li}68}$. The true breakup temperature is likely to be
underestimated by this observable but its trend with incident 
energy may be preserved. Therefore, at this stage, no attempt has been 
made to derive corrections, and the so-called apparent temperatures,
labelled with the subscript 0 in the above expressions, are presented in 
Table~\ref{tab:table2} and Fig.~\ref{fig:fig5}.
The differences of the energy intervals of the fragment detection 
\cite{boga74,boga80} and the systematic errors associated 
with the isotope identification \cite{boga82} are taken into account.

The deduced values of $T_{\rm HeLi,0}$ and $T_{{\rm He}68/{\rm Li}68,0}$ 
of about 4 to 5 MeV are in the range typical for reaction processes 
near the onset of multi-fragment emissions 
\cite{poch95,tsang97,kwiat98,haug2000}. 
The values obtained for $T_{{\rm He}46/{\rm Li}68,0}$ are lower by 
1 MeV or more, as expected.
Most of the temperatures, within errors, are about equal for the 
corresponding pairs of reactions. Larger differences, as e.g. of 
$T_{\rm HeLi,0}$ for protons of 660 MeV, reflect similarly prominent 
deviations from isoscaling for some of the isotopes involved 
(cf. Fig.~\ref{fig:fig1}). 

The trend with incident energy exhibited by 
$T_{{\rm He}46/{\rm Li}68,0}$ is found to follow very closely
that of the inverse of the scaling parameters 
$\alpha$ and $\beta_{t3}$ (Fig.~\ref{fig:fig5}). 
This suggests that the gradual
flattening of the slopes of the isoscaling curves,
as the projectile energy increases, is indeed caused
by a rising mean temperature. A variation of isotopic observables 
associated with a temperature change has recently 
been reported for the fragmentation of $^{28}$Si projectiles in
collisions with $^{112,124}$Sn targets at 30 and 50 MeV per nucleon
\cite{vesel01}.

\section{\label{sec:smmin}SMM interpretation}

The statistical multifragmentation model (SMM) 
is based upon the assumption of statistical equilibrium at a
low-density freeze-out stage \cite{bond95}. 
All breakup channels (partitions) composed of nucleons and
excited fragments are considered and the conservation of mass, charge, 
momentum and energy is taken into account.
The formation of a compound nucleus is included as one of the channels. 
In the microcanonical 
treatment the statistical weight of decay channel j is given by 
$W_{\rm j} \propto exp~S_{\rm j}$, where $S_{\rm j}$ is the entropy of 
the system in channel j which is a function of the excitation 
energy $E_{\rm x}$, mass number $A_{\rm s}$, charge $Z_{\rm s}$ 
and other parameters of the source.
In the standard version of the model, the Coulomb interaction between
the fragments is treated in the Wigner--Seitz approximation.
Different breakup partitions are sampled according to their statistical
weights uniformly in the phase space.
After breakup,
the fragments propagate independently in their mutual Coulomb field and
undergo secondary decays.
The deexcitation of the hot primary fragments
proceeds via evaporation, fission, or Fermi-breakup \cite{botvina87}.

\subsection{\label{sec:liqui}Liquid-drop description of primary fragments}

An important difference of the SMM from other statistical models, 
e.g. QSM \cite{hahn88} or the Berlin statistical multifragmentation model
\cite{gross90,snepp94}, is the treatment of the hot fragments at the 
freeze-out density. In the SMM 
light fragments with mass number $A\le 4$ are considered as stable
particles ("nuclear gas") with masses and spins taken from the nuclear 
tables. Only translational degrees of freedom of these particles 
contribute to the entropy of the system. 
Fragments with $A > 4$ are treated as heated nuclear liquid  drops, 
and their 
individual free energies $F_{AZ}$ are parameterized as a sum of the bulk, 
surface, Coulomb and symmetry energy contributions 
\begin{equation}
F_{AZ}=F^{B}_{AZ}+F^{S}_{AZ}+E^{C}_{AZ}+E^{sym}_{AZ}.
\end{equation}
The standard expressions \cite{bond95} for these terms are:
$F^{B}_{AZ}=(-W_0-T^2/\epsilon_0)A$, where the
parameter $\epsilon_0$ is related to the level density, and 
$W_0 = 16$~MeV is the binding energy of infinite nuclear matter; 
$F^{S}_{AZ}=B_0A^{2/3}(\frac{T^2_c-T^2}{T^2_c+T^2})^{5/4}$, where
$B_0=18$~MeV is the surface coefficient, and $T_c=18$~MeV is the critical 
temperature of infinite nuclear matter; $E^{C}_{AZ}=cZ^2/A^{1/3}$, where 
$c$ is the Coulomb parameter obtained in the Wigner-Seitz 
approximation, $c=(3/5)(e^2/r_0)(1-(\rho/\rho_0)^{1/3})$, with the charge 
unit $e$ and $r_0$=1.17 fm; $E^{sym}_{AZ}=\gamma (A-2Z)^2/A$, where 
$\gamma = 25$~MeV is the symmetry energy parameter. 

These parameters are those of the Bethe-Weizs\"acker formula and correspond 
to the assumption of isolated fragments with normal density in the 
freeze-out configuration, an assumption found to be quite successful in 
many applications. It is to be expected, however, that in a 
more realistic treatment primary fragments will have to be considered
not only excited but also expanded and still subject to a residual nuclear 
interaction between them.
These effects can be accounted for in the fragment 
free energies by changing the corresponding liquid-drop parameters,
provided such modifications are also indicated by the experimental data. 
In the following, it will be shown that, for the symmetry energy,
this information may be obtained from the isoscaling phenomenon. 

\subsection{\label{sec:grand}Grand canonical approximation}

In the grand canonical approximation, first developed in 
Ref.~\cite{botvina85}, the mean
multiplicity of a fragment with mass number $A$ and charge $Z$ is
given by
\begin{equation} \label{eq:naz}
\langle N_{AZ}\rangle
=g_{AZ}\frac{V_{f}}{\lambda_{T}^{3}}A^{3/2}\exp\left[-\frac{1}{T}
\left(F_{AZ}(T,\rho)-\mu A-\nu Z\right)\right]
\end{equation}
where 
$g_{AZ}$ is the degeneracy factor of the fragment, 
$\lambda_{T}$ is the nucleon thermal wavelength, $V_f$ is the "free" volume, 
and $\mu$ and $\nu$ are the chemical potentials responsible for the 
mass and charge conservation in the system, respectively \cite{bond95}. 
It follows immediately that, for two systems 1 and 2 with different 
total mass and charge but with the same temperature and density, 
the ratio of fragment yields produced in these systems is given 
by Eq.~(\ref{eq:scalab}) with parameters 
$\alpha=(\mu_1-\mu_2)/T$ and $\beta=((\mu_1-\mu_2)+(\nu_1-\nu_2))/T$. 
Isoscaling arises very naturally in the SMM. 

\begin{figure}
     \epsfysize=9.0cm
     \centerline{\epsffile{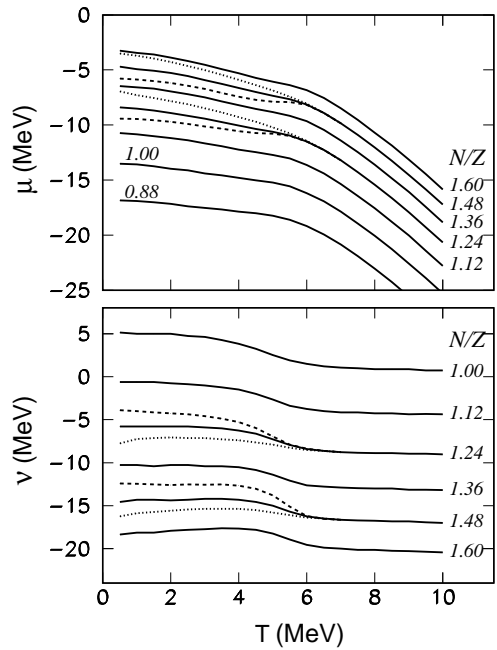}}
\caption{\label{fig:fig6} 
Results of SMM calculations in the grand canonical approximation
for the chemical
potentials $\mu$ (top) and $\nu$ (bottom) as a function of the
temperature of systems with different $N/Z$ ratios as indicated and
with sizes $Z$ = 50 (solid lines), $Z$ = 100 (dashed), and $Z$ = 25 
(dotted). The density is $\rho = \rho_0$/3.
}
\end{figure}

Calculated chemical potentials for systems with different mass and
$N/Z$ ratio as a function of the temperature are shown in
Fig.~\ref{fig:fig6}. 
A freeze-out density $\rho /\rho_0$ = 1/3 has been chosen but, 
as apparent from the QSM calculations (Fig.~\ref{fig:fig4}), other 
densities lead to a similar behavior of the chemical potentials. 
Furthermore, corresponding to the excluded
volume approximation \cite{bond95}, a fixed "free" volume 
$V_f=2V_0$ ($V_0$ is the volume of the system at normal density) 
has been used instead of a multiplicity-dependent volume.
For other parameters of the model 
their standard values were chosen, see e.g. Ref. \cite{botv95}.

The potential $\mu$ decreases with the temperature which has the simple 
physical meaning that the average size (mass number) of the produced 
fragments decreases. However, two regions with different rates of the 
change in fragment mass can be discerned. 
At low temperature, the rate is small, 
especially for large systems. Here the corresponding mass distribution 
is of the so-called "U-shape", with a compound-like fragment still 
dominating in the system. 
At temperatures near 5 to 6~MeV the rate increases 
rapidly. At this point, the "U-shape" disappears and the system 
disintegrates into many fragments with an approximately exponential
mass distribution. 

The behavior of the chemical potential $\nu$ is particularly interesting.
As shown in Ref.~\cite{botvina87}, the average charge 
$\langle Z_{A} \rangle$ of fragments with mass $A$ can be written as 
\begin{equation} \label{eq:za}
\langle Z_{A} \rangle \simeq \frac{(4\gamma+\nu)A}{8\gamma +2cA^{2/3}}.
\end{equation}
The chemical potential $\nu$ is, therefore, directly connected with 
the isospin of the produced fragments.
In grand-canonical models, as e.g. in the QSM \cite{hahn88}, sometimes
the chemical potentials $\mu_n$ and $\mu_p$, responsible for the 
conservation of the total numbers of neutrons and protons, are used. If
expressed in terms of the SMM potentials, they are
$\mu_n=\mu$ and $\mu_p=\mu+\nu$, with the consequence that not only
$\mu_n$ but also $\mu_p$ will decrease with $T$; the mean numbers of both, 
neutrons and protons, in the produced fragments decreases
as their mass decreases. By using the potentials $\mu$ and $\nu$ the 
variations of the average mass and of the isotopic composition
are separated. 

The potential $\nu$ is nearly constant at both low and high 
temperature. These limits correspond to the isospin of fragments produced at 
the "liquid" and "gas" phases. There is a relatively fast transition between 
these limits at a temperature of 5 to 6~MeV, leading to a growing 
neutron content of light fragments as the "U-shape" disappears. 
This evolution of the fragment isospin has been confirmed by microcanonical 
calculations \cite{botvina01}. It is apparent from Fig.~\ref{fig:fig6} 
that the effect 
is more pronounced for larger systems and that its relative magnitude 
depends weakly on the overall $N/Z$ ratio.

For systems with different mass but with the same $N/Z$ ratio,
the chemical potentials differ only in the "U-shape" region
at low temperatures. At high temperature, the chemical potentials coincide
which leads to a total scaling of the fragment yields for systems of all 
sizes. This is equivalent to what is obtained by applying the 
grand canonical ensemble for the region of abundant multifragmentation. 
Examples of fragment charge distributions that are independent of the 
system size have been presented recently \cite{rivet98}.

\subsection{\label{sec:chemp}Chemical potentials}

\begin{figure}
     \epsfysize=8.5cm
     \centerline{\epsffile{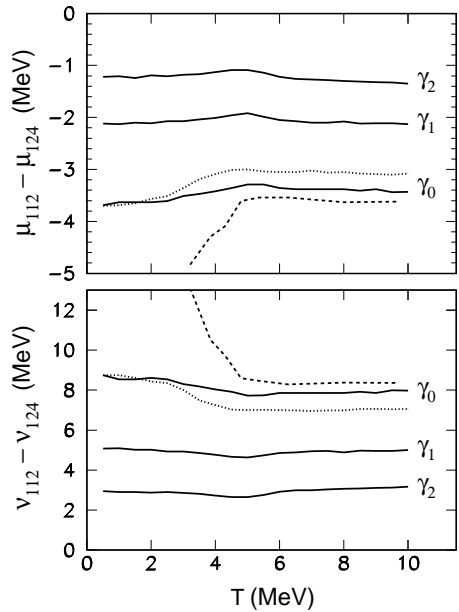}}
\caption{\label{fig:fig7} 
Differences of the chemical potentials $\mu$ (top) and $\nu$ (bottom)
for the $^{112}$Sn and $^{124}$Sn systems as a function of the temperature
and for symmetry term coefficients
$\gamma_0$ = 25~MeV,
$\gamma_1$ = 14.4~MeV and $\gamma_2$ = 8.3~MeV.
Solid and dotted lines represent grand
canonical calculations for $\rho=\rho_0$/3 and $\rho=\rho_0$/6, 
respectively.
The dashed lines are microcanonical Markov-chain calculations for
$\rho=\rho_0$/3.
}
\end{figure}

It is the difference of the chemical potentials of 
systems with different $N/Z$ ratios that is directly connected with the 
isoscaling phenomenon. Results of calculations for the two tin isotopes
are shown in Fig.~\ref{fig:fig7}. 
Despite of a considerable variation of the individual 
potentials, their differences $\Delta \mu = \mu_{112}-\mu_{124}$ and 
$\Delta \nu = \nu_{112}-\nu_{124}$ change only slightly as a function
of the temperature. For the lower density $\rho /\rho_0$ = 1/6,
an increased modulation is observed but, overall, the potential differences
remain remarkably stable in the most important temperature region. 
This means that a variation of the scaling parameters $\beta_{t3}$
or $\alpha = \Delta \mu/T$ is connected with a temperature change as 
suggested by the comparisons shown in Fig.~\ref{fig:fig5}. 

Calculations were also performed 
with the Markov-chain version of the SMM with parameters identical to those
used for the grand-canonical calculations. The Markov-chain model
is a completely microcanonical  
approach which exactly conserves mass, charge, energy, and linear and 
angular momentum \cite{botvina01}. It was used to calculate the ratios of 
isotopes produced by the two different sources and the microcanonical 
temperature $T_{\rm micr}$ of the system, 
where $T_{\rm micr}$ is taken as the mean 
temperature of the generated partitions at a fixed excitation energy.
From these quantities
effective potential differences $\Delta \mu$ and $\Delta \nu$
were determined according to 
\begin{equation}
\Delta \mu=T_{\rm micr}*ln(\frac{Y_1(A,Z)Y_2(A+1,Z)}{Y_2(A,Z)Y_1(A+1,Z)}),
\end{equation}
\begin{equation}
\Delta \nu=T_{\rm micr}*ln(\frac{Y_1(A,Z)Y_2(A,Z+1)}{Y_2(A,Z)Y_1(A,Z+1)}).
\end{equation}
Here only light fragments with $Z$ from 3 to 6 were used, similar to the 
experimental case. 
These potentials are very close to the grand-canonical 
results at temperatures $T > 5$~MeV which are of
relevance for the production of fragments. 
The remaining small difference may arise from the need of 
using slightly different ensembles in the two types of calculations. 
At low temperatures the results diverge, indicating that here 
the exact conservation of mass, charge and energy is essential
(cf. Fig. 5.6 in Ref. \cite{bond95}). 
A small difference to the results reported in Ref. \cite{tsang01a}
exists insofar as the variations of $\Delta \mu_{\rm n}$ and $\Delta 
\mu_{\rm p}$ at high temperature, obtained there 
in the canonical approximation,
are not reproduced by the present calculations.
It may be a consequence of the constraint of the energy conservation,
of another method of calculating chemical potentials, or 
of other differences of the used model versions.
The grand-canonical QSM calculations predict 
negligible variations of the potential differences with temperature.

The symmetry term in the binding energy strongly influences the potential
differences $\Delta \mu$ and $\Delta \nu$. This is illustrated 
in Fig.~\ref{fig:fig7} 
with examples obtained for values $\gamma_1$ = 14.4 MeV and
$\gamma_2$ = 8.3 MeV,
both smaller than the standard value $\gamma_0$ = 25 MeV.
The effect is significant and, in particular, larger than the 
variations associated with the choice of ensembles or with the 
choice of the density
and, therefore, should be observable. 

As the comparison shows, 
the grand-canonical approach is applicable only (i) if 
the average mass of the largest fragment of a partition is considerably
less than the total size of the system, and (ii) if the extra energy 
necessary for the production of an additional fragment 
is small compared to the available thermal energy.
However, because the difference of the chemical potentials is nearly 
constant in the full temperature range, the values obtained in the
grand-canonical approximation at low temperatures may be extrapolated
to high temperatures and applied in the multifragmentation region.
In the low temperature limit $T\rightarrow 0$, analytical formulae 
for $\Delta \mu$ and $\Delta \nu$ can be derived.
Here only channels including a
compound-like nucleus with $A \approx A_0$ and $Z \approx Z_0$ will exist,
where $A_0$ and $Z_0$ denote the mass and atomic number of the system. 
Mathematically, it is required that the numerator 
under the exponent in Eq.~(\ref{eq:naz}) approaches zero, i.e. 
\begin{equation}
F_{A_0Z_0}(T\rightarrow 0)=\mu A_0+\nu Z_0,
\label{eq:faz}
\end{equation}
which is equivalent to the thermodynamical potential of the compound 
nucleus being zero. From Eq.~(\ref{eq:za}), with the same approximations,
the potential
\begin{equation}
\nu \simeq \frac{Z_0}{A_0}(8\gamma +2cA^{2/3}_0)-4\gamma
\end{equation}
can be obtained. Inserting this expression into
Eq.~(\ref{eq:faz}) yields the chemical potential
\begin{equation}
\mu \simeq -W_0+\frac{B_0}{A_0^{1/3}}-c\frac{Z_0^2}{A_0^{4/3}}+
    \gamma(1-(\frac{2Z_0}{A_0})^2).
\label{eq:mu}
\end{equation}
The terms small compared to the bulk terms can be safely disregarded
(the errors are below 3\% for the $^{112,124}$Sn isotopes considered 
here). This leads to
\begin{eqnarray} \label{eq:dmunu}
\Delta \mu = \mu_1 - \mu_2 \approx -4\gamma 
(\frac{Z_{1}^2}{A_{1}^2}-\frac{Z_{2}^2}{A_{2}^2}),\nonumber\\
\Delta \nu = \nu_1 - \nu_2 \approx 8\gamma 
(\frac{Z_{1}}{A_{1}}-\frac{Z_{2}}{A_{2}}),
\end{eqnarray}
where $Z_{1}$,$A_{1}$ and $Z_{2}$,$A_{2}$ are the charges and mass 
numbers of the two systems. The potential differences 
depend essentially only on the coefficient $\gamma$ of the symmetry term 
and on the isotopic compositions. 

The values of the chemical potentials deduced in this limit are 
close to the separation energies of nucleons, apart from the difference in
sign (see also Ref. \cite{tsang01a}). 
For example, the neutron separation energy $s_n$ 
in the liquid-drop approximation is given by
\begin{equation}
s_n\approx W_0-\gamma(1-(\frac{2Z_0}{A_0})^2)-\frac{2B_0}{3A_0^{1/3}}
+\frac{e^2Z_0^2}{5r_{0}A_0^{4/3}}. 
\end{equation}
The surface and Coulomb terms in this expression appear with different 
coefficients than in Eq.~(\ref{eq:mu}) but are, again, usually 
small in comparison to the dominating bulk (volume and symmetry) terms.
As expected from the definitions of the chemical potential and the 
separation energy, this correspondence must be exact in the thermodynamical 
limit. 

From Eqs.~(\ref{eq:dmunu}) another interesting relation can be deduced: 
\begin{equation} \label{eq:cmunu}
\Delta\mu\approx-\frac{\Delta\nu}{2}(\frac{Z_{1}}{A_{1}}+\frac{Z_{2}}{A_{2}}).
\end{equation}
It implies $|\Delta \nu|>|2 \Delta \mu|$ or, equivalently,
$|\Delta \mu_p|>|\Delta \mu_n|$, for the usually considered systems with
$A > 2Z$. 
Although the effects of secondary deexcitation are important 
(see below) this inequality is reflected by the observed scaling 
parameters. The magnitude of $\beta$ exceeds that of $\alpha$ in 
all reactions discussed in Ref.~\cite{tsang01} and, on average, also 
in the reactions presented here (Table~\ref{tab:table1}).

\subsection{\label{sec:width}Fragment distribution widths}

%There is a simple explanation for the SMM scaling appropriate for 
%finite systems.
There is a simple physical 
explanation within the SMM why isoscaling should appear
in finite systems.
Charge distributions of fragments with fixed mass numbers $A$, as well 
as mass distributions for fixed $Z$, are approximately Gaussian with 
average values and variances which are connected with the temperature,
the symmetry coefficient, and other parameters \cite{botvina85}. 
With a Gaussian distribution for an observable $X$ 
(mass number or charge), 
$Y(X)\propto exp(-(X-\langle X\rangle)^2/2\sigma^2)$, the ratio of
this observable for two different systems is given by
\begin{equation} \label{eq:expon}
\frac{Y_1(X)}{Y_2(X)}=exp(
-\frac{X^2}{2}(\frac{1}{\sigma_1^2}-\frac{1}{\sigma_2^2})
+X(\frac{X_1}{\sigma_1^2}-\frac{X_2}{\sigma_2^2})+const),
\end{equation}
where $X_1, X_2$ and $\sigma_1, \sigma_2$ are the mean values and variances 
for the two systems. The mean values depend on the total 
mass and charge of the systems, e.g. via the chemical potentials in the 
grand canonical approximation (Eq.~(\ref{eq:za})), while 
the variances depend mainly on the physical conditions reached,
the temperature, the density and possibly other variables. For example, 
the charge variance $\sigma_Z\approx \sqrt(AT/8\gamma)$ obtained for
fragments with a given mass number $A$ in Ref. \cite{botvina85} 
is only a function of the temperature and of the symmetry term coefficient
since the Coulomb contribution is very small.
If these physical conditions are the same, i.e. $\sigma_1 = \sigma_2$, 
the exponential scaling for the ratio follows from Eq.~(\ref{eq:expon}).
Furthermore, by using Eqs.~(\ref{eq:za}) and (\ref{eq:dmunu}) for $X = Z$,
the approximate relation $\beta = \Delta \nu /T$ is again obtained, as in 
the usual grand-canonics.

\begin{figure}
     \epsfysize=8.0cm
     \centerline{\epsffile{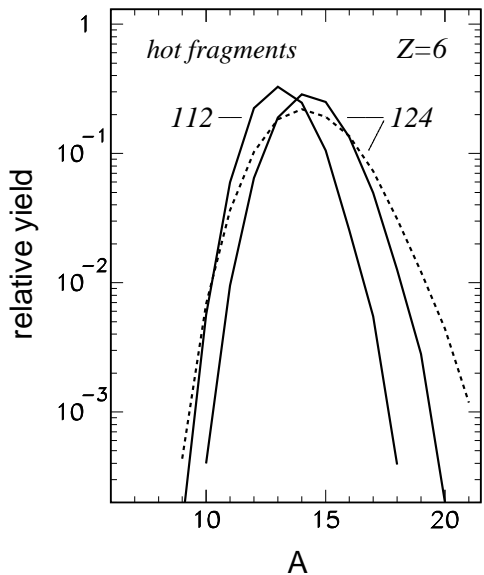}}
\caption{\label{fig:fig8} 
Mass distributions of primary hot fragments with $Z$ = 6 produced at
freeze-out by $^{112}$Sn and $^{124}$Sn systems, as obtained from
Markov-chain calculations for $E_{\rm x}/A$ = 5~MeV and $\rho=\rho_0$/3
(the corresponding microcanonical
temperature is $T_{\rm micr} \approx$ 5.3~MeV).
The symmetry coefficients $\gamma$ = 25~MeV (solid lines)
and $\gamma$ = 14.4~MeV (dashed line) were used.
}
\end{figure}

The Gaussian distributions 
obtained in the grand-canonical approximation are reproduced
by the Markov-chain SMM calculations (Fig.~\ref{fig:fig8}). The
mass distributions of fragments with $Z=6$ emitted by $^{112}$Sn and 
$^{124}$Sn with $E_{\rm x}/A$ = 5 MeV are shifted with respect to each other
because the $N/Z$ ratios of the sources are different. Scaling will result,
and the value of the scaling coefficient is determined 
by both, the shift, i.e. the difference in the mean masses, and the 
width of the distributions. The width, in turn, is influenced by the
symmetry coefficient; with a reduced coefficient $\gamma$ the mass
distribution widens considerably (Fig.~\ref{fig:fig8}). 
Thus, if the temperature is known
the symmetry coefficient can, in principle, 
be determined using the distributions.

\begin{figure}
     \epsfysize=8.0cm
     \centerline{\epsffile{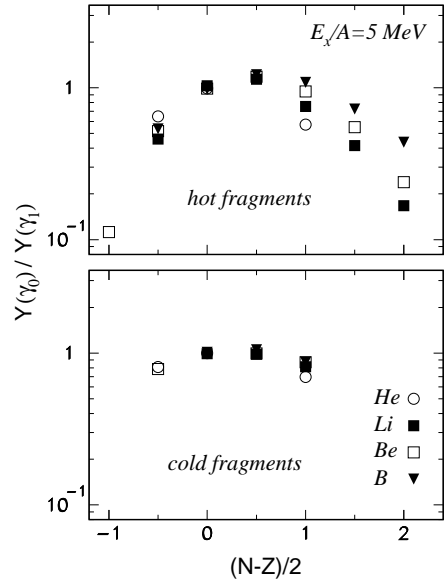}}
\caption{\label{fig:fig9} 
Ratios of isotopic yields calculated for different symmetry coefficients
$\gamma_0$ = 25~MeV and $\gamma_1$ = 14.4~MeV for the breakup of a
$^{112}$Sn source with $E_{\rm x}/A$~=~5~MeV at a density $\rho=\rho_0$/3.
The top and bottom panels give the ratios for hot and cold fragments,
respectively.
}
\end{figure}

The calculations indicate that the secondary deexcitation reduces
both, the differences between the mean values of the distributions
and the magnitude of the variances,
thereby attaching a considerable uncertainty to this method. 
However, the sensitivity to the symmetry term coefficient survives the 
deexcitation stage. This is illustrated in Fig.~\ref{fig:fig9} which shows 
the SMM predictions for the ratios of isotopic yields that are 
obtained for the same thermal source with different 
coefficients $\gamma$.
The characteristic bell shape of the distributions reflects the quadratic 
term of Eq.~(\ref{eq:expon}) which dominates in this case when
$\sigma_1 \neq \sigma_2$. 

\subsection{\label{sec:secon}Secondary deexcitation of fragments}

In the SMM the secondary 
deexcitation of large fragments with $A>16$ is described with 
Weisskopf type evaporation and Bohr-Wheeler type fission models while 
the decay of small fragments is treated with a Fermi-breakup 
model \cite{bond95,botvina87}. In this model all ground and 
nucleon-stable excited states of light fragments are taken into 
account and the population probabilities of these states are calculated
according to the available phase space. The model thus simulates a 
simultaneous breakup microcanonically. This procedure is expected
to reliably describe a decay that happens at short time scales 
after the freeze-out if the excitation energy of the primary 
fragments is high, of the order of 2-3~AMeV or higher.

\begin{figure}
     \epsfysize=10.0cm
     \centerline{\epsffile{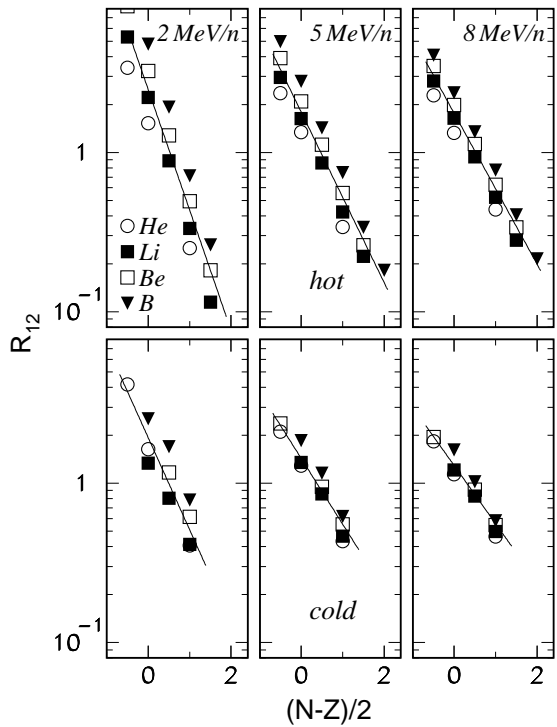}}
\caption{\label{fig:fig10} 
Ratios of isotope yields produced at the breakup of $^{112}$Sn and
$^{124}$Sn sources from Markov-chain SMM calculations for three
excitation energies $E_{\rm x}/A$~=~2, 5
and 8~MeV and a density $\rho=\rho_0$/3.
The top and bottom panels are for hot and cold fragments, respectively.
The solid lines correspond to the logarithmic slope parameters
$\beta_{t3}$ given in Table~\protect\ref{tab:table3}.
}
\end{figure}

The ratios of light element yields (2 $\le Z \le$ 5)
calculated with the Markov-chain 
SMM are shown in Fig.~\ref{fig:fig10} for hot fragments produced at breakup 
and for cold fragments after the sequential decay. The exponential scaling
with isospin is observed for both cases but with scaling 
coefficients that are systematically smaller for the final cold 
fragments (Table~\ref{tab:table3}). 
On more general grounds, 
it is expected that the scaling property is preserved
because the excitation energies per nucleon are similar
for all fragments, so that their relative nucleon content will decrease 
in a similar way. The secondary deexcitation has a trend, however, 
to populate the $\beta$-stable region which may reduce the shift between 
the mass distributions and also reduce their widths. 
A modification of the scaling coefficients
is thus expected even though these two effects may partially 
compensate each other. In this respect, different isotopes can behave 
differently. The predicted reduction of the mass 
widths is typically 30\% for boron isotopes, i.e. significant as expected, 
but is practically negligible for the lithium isotopes.

\begin{table}
\caption{\label{tab:table3}
Parameters obtained from fitting the yield ratios
of isotopes with 2 $\le Z \le$ 5 as
calculated with the Markov-chain SMM 
for excitation energies $E_{\rm x}/A$ = 2, 5 and 8 MeV 
with the scaling functions given in 
Eqs.~(\protect\ref{eq:scalab}) and (\protect\ref{eq:scalboga}). 
Uncertainties are of the order of 0.01 to 0.02.
}
\begin{ruledtabular}
\begin{tabular}{c c c c c}
%\hline\hline
$E_{\rm x}/A$ (MeV) & & $\beta_{t3}$ & $\alpha$ & $\beta$ \\
\hline
2 & hot & 1.74 & 0.93 & -1.31 \\
5 & " & 1.23 & 0.67 & -0.91 \\
8 & " & 1.09 & 0.57 & -0.77 \\
\hline
2 & cold & 1.33 & 0.69 & -0.82 \\
5 & " & 0.95 & 0.52 & -0.63 \\
8 & " & 0.84 & 0.45 & -0.55 \\
%\hline\hline
\end{tabular}
\end{ruledtabular}
\end{table}

According to the calculations, the coefficients $\beta_{t3}$ and
$\alpha$ are reduced to, on average, 77\% of their values
by the secondary decay (Table~\ref{tab:table3}). 
The coefficient  $\beta$ is more strongly reduced to about 
60\% of its value at 2~AMeV and to about 70\% at 8~AMeV.
However, since the coefficients are decreasing with the energy, the
absolute magnitude of this reduction decreases also. Moreover,
additional calculations showed that the secondary-decay effect decreases
considerably if the coefficients themselves become smaller, e.g., a primary
$\alpha \approx$ 0.46 is reduced to $\approx$ 0.44.
In this respect, we confirm the conclusions of Ref.~\cite{tsang01a} 
regarding the minute variation of the $\alpha$ parameter, 
but provided that the value of the initial $\alpha$ is relatively small.
The calculations of Ref.~\cite{tsang01a} were done for $E_{\rm x}/A$ = 6~MeV
with several statistical models, including other versions of the SMM.
In some of these calculations, the decay procedure
is based on a sequential emission of particles from primary
fragments, following the tabulated branching ratios and a Weisskopf scheme.
This seems adequate for a later deexcitation stage with
isolated fragments at relatively
low excitation energy and without the influence of a common Coulomb
field and without a residual nuclear interaction which
can modify fragment properties including the branchings.
The obtained modifications of the scaling parameter
$\alpha$ do not exceed the order of 5\% whereas $\beta$ is
reduced more strongly, similar to the present case.

The primary values of the scaling parameter $\alpha \approx$ 0.4 to
0.45 reported in Ref. \cite{tsang01a} for the $^{112,124}$Sn systems
are smaller than
the corresponding values given in Table~\ref{tab:table3}. 
This, apparently, reflects
significant differences between different versions and 
parameterizations of, in principle,
the same model. They are of the same order as potential effects of the
symmetry term that are to be studied.
This emphasizes the need for exclusive analyses of experimental data which 
should constrain the model parameters.
The two secondary deexcitation procedures
should be considered as partly complementary, and the range of the
differences of the obtained results may characterize the reliability
of treating secondary decays with model calculations.
These corrections are essential \cite{zubkov97},
but it will be important to reduce the uncertainties. 
Experimental methods,
e.g. based on correlation techniques \cite{marie98}, may prove
very useful for this purpose.

\subsection{\label{sec:inter}Interpretation of the data}

The deduced relations will now be used for the interpretation of the 
experimental data. We will concentrate on the two reactions initiated
by the projectiles with the highest energies, protons of 6.7 GeV and 
$\alpha$ particles of 15.3 GeV, for which the contributions from 
instantaneous 
breakups into multifragment channels should be enhanced in comparison to 
the other cases. The inclusive nature of the measurements, nevertheless,
presents an inherent difficulty since a wide range of excitation energies 
is covered by the fragment emitting sources.

The $^{112,124}$Sn targets used in these experiments were isotopically
enriched to 
81.7\% and 96.6\%, respectively \cite{boga80,boga82}. The effects of the 
impurities, known to be distributed approximately as the natural 
abundances of tin isotopes, 
have to be taken into account in a quantitative analysis.
Corrections were estimated by assuming Gaussian mass distributions for the
produced fragments, centered around mean values that vary linearly with the 
mass number of the considered tin isotopes. It was found that, for the
specific enrichments of the used targets and for scaling coefficients
$\alpha$ in the range 0.3 to 0.6, the impurities cause a reduction of the 
measured $\alpha$ by 10\% to 15\%.

The analytical expressions for the differences of the chemical potentials,
derived in the grand-canonical approximation 
(Eqs.~(\ref{eq:dmunu})), depend only on $\gamma$ and the 
isotopic composition of the sources. In the case of $\Delta \mu$, the 
difference of the squared $Z/A$ values is required which is
found to be the same within a few percent, independently of whether
it is evaluated for the original targets $^{112,124}$Sn or for
the excited systems as predicted by the INC
calculations (Fig.~\ref{fig:fig3}, Section~\ref{sec:dynam}).
For the original targets it amounts to 
$(Z_1/A_1)^2 - (Z_2/A_2)^2$ = 0.0367, leading to 
$\gamma = \Delta \mu /0.147$.
To obtain an experimental value of 
$\Delta \mu = \alpha \cdot T$ (Section~\ref{sec:grand}),
the mean values of the scaling coefficient $\alpha$ and of the isotope 
temperature $T_{{\rm He}68/{\rm Li}68}$ for the p(6.7 GeV) and 
$\alpha$(15.3 GeV) reactions are used, after applying corrections.
A measured $\langle \alpha \rangle$ = 0.365 is obtained from 
Table~\ref{tab:table1},
corresponding to 0.417 for isotopically pure targets, 
and the effect of the secondary deexcitation is assumed to be 23\%,
as suggested by the Markov-chain calculations (Table~\ref{tab:table3}), 
thus leading to a primary $\alpha$ = 0.542.

The predictions of the QSM \cite{hahn88} are used for the correction
of the temperature. It does not significantly depend on the assumed density
but it is large, as expected. The mean apparent temperature
$T_{{\rm He}68/{\rm Li}68,0}$ = 4.35 MeV (Table~\ref{tab:table2})
corresponds to a breakup temperature $T$ = 6.2 MeV in this model.
The results obtained with these inputs are 
$\Delta \mu$~=~3.36 MeV and $\gamma$ = 22.8 MeV, a symmetry coefficient
slightly but not significantly smaller than the adopted standard value
of 25 MeV.  

For the interpretation of the isoscaling coefficient in the microcanonical 
limit the excitation energy needs to be specified. Exclusive data 
for hadron induced reactions on Au targets indicate that 
fragments will be emitted if energies exceeding $\approx$ 400 MeV, 
corresponding to $E_{\rm x}/A \approx$ 2 MeV, are deposited 
by the projectile \cite{beau99,beau00}. 
Since the cross sections decrease and the fragment emission probabilities 
increase with excitation energy, a rather wide distribution results.
For the $\pi^-$ projectiles of 8 GeV/c studied by the ISiS collaboration
this distribution extends from below 3 to above 8 MeV per nucleon 
with a weighted mean value of 
$E_{\rm x}/A \approx$ 5 MeV \cite{beau99,beau00}.
A similar or, because of the lighter targets, a slightly higher value 
may be expected for the case of protons of 6.7 GeV on $^{112,124}$Sn.
The INC calculations for this reaction, 
again weighted by the fragment production cross section, 
predict an average excitation 
energy $E_{\rm x}/A$ = 6.2 MeV. 
With this interval 5.0 MeV to 6.2 MeV per nucleon for the excitation 
energy, and with the assumption that $\alpha \propto \gamma$ as in the 
grand-canonical approximation, values between $\gamma$ = 21.4 MeV 
and 22.6 MeV are obtained from the comparison of the measured 
$\alpha$ = 0.39 ($\alpha$ = 0.45 for pure targets) with the predictions 
given in Table~\ref{tab:table3} for which 
the standard value $\gamma$ = 25 MeV was used.
If $\langle E_{\rm x}/A \rangle$ =~8~MeV is considered as realistic for
$\alpha$(15.3 GeV) a similar symmetry 
coefficient $\gamma$ = 21.6 MeV will result.

Towards the lower projectile energies, the isoscaling
coefficient $\alpha$ increases 
up to 0.53, corresponding to 0.61 for pure targets,
which is still lower than the SMM predictions
for small excitation energies (Table~\ref{tab:table3}). 
With the INC result
$\langle E_{\rm x}/A \rangle$ = 2.7 MeV for protons of 660 MeV,
the interpolated prediction is $\alpha$ = 0.65, and 
$\gamma$ = 23.3 MeV is obtained from the comparison
with the measured value. It thus seems that, for the reactions studied 
here, the deduced values of $\gamma$ fall consistently into the range of 21 
to about 23 MeV, with no significant dependence on the energy. In this 
respect, however, it has to be considered that the constraint of energy 
conservation in the microcanonical calculations may lead to
unrealistically narrow widths of the 
isotope distributions at low excitation energies. This would cause an
overprediction of the scaling coefficients and a deduced $\gamma$ that is 
too low. This effect will bring $\gamma$ even closer to the standard value
for the reactions at lower incident energies which 
primarily proceed via the formation of excited compound nuclei. 

\section{\label{sec:summa}Summary and conclusions}

In the first part of this paper,
the existence of isoscaling for reactions induced by relativistic 
light particles was demonstrated. 
The deduced exponents vary smoothly with the incident 
energy. Their trends, apparently, extend beyond the range studied here
to low-energy 
projectiles as, e.g., $\alpha$ particles of 200 MeV for which isoscaling 
parameters were reported in Ref. \cite{tsang01}.
The values obtained for protons of 6.7 GeV and
$\alpha$ projectiles of 15.3 GeV are close to those for
central $^{112,124}$Sn + $^{112,124}$Sn reactions at 50 MeV per nucleon
given in the same reference. The observation of $t_3$ scaling was 
illustrated and discussed. As a function of the projectile energy,
a very similar variation of the inverse scaling parameters and of the 
isotope temperature $T_{{\rm He}46/{\rm Li}68}$ was observed.

In the second part, a statistical formalism for the interpretation of the 
isoscaling phenomenon was developed. Analytical expressions were derived in 
the grand-canonical approximation and their validity and applicability 
illustrated. Results of calculations in the grand-canonical approximation
and with the microcanonical 
Markov-chain version of the SMM were presented and the
connection with the symmetry term of the fragment binding energy was 
established. It was found that the difference 
of the chemical potentials for the two 
isotopically different systems does not depend on the temperature. For the 
Markov-chain calculations, this conclusion is valid for temperatures
$T \ge$ 5 MeV, the range of relevance for multifragment processes. The 
invariance of $\Delta \mu$ with temperature is consistent with the 
interpretation that the observed variation of the scaling parameters
is caused by a change in temperature, 
as suggested by the temperature measurement. 

In the last part (Section~\ref{sec:inter}), 
an attempt was made to deduce values for 
the symmetry-energy coefficient $\gamma$ from the experimental data. 
The analytical formulae derived in the grand-canonical limit of the SMM and 
the results of the microcanonical calculations were used and very 
similar values in the range $\gamma = 22.5 \pm 1$ MeV were obtained.
Besides the scaling coefficient, experimental values for either the 
breakup temperature in the grand-canonical or for the excitation energy 
in the microcanonical approach were required. In the latter case, estimates 
obtained for similar reactions and from INC calculations were used.

We estimate the uncertainties of the methods, in particular 
the errors associated with the determinations of the breakup temperature
or of the excitation energy for the microcanonical method, 
to be at least of the same order as the deviations of the results from the 
standard value $\gamma$ = 25 MeV. The sequential decay 
corrections are substantial and, e.g., in the grand-canonical case
are required twice, for the scaling coefficient and for the temperature.
The present results, therefore, do not contradict the assumptions made in 
the statistical multifragmentation model in using standard liquid-drop
parameters for describing the nascent fragments at the breakup stage.

A problem associated with the present data is the wide range
of excitation energies over which an average is taken in the inclusive
measurements. Smaller variations may be smeared out. 
For these reasons, the presented analysis is primarily intended to serve 
as an example of how to extract the symmetry-energy coefficient $\gamma$ 
from the experimental data.
It is, nevertheless, of interest that the
obtained result for fragmentation reactions
induced by relativistic light projectiles has a tendency to be smaller
than the conventional value of 25 MeV. A reduction with increasing 
energy may even be suggested by the microcanonical analysis.
Provided it can be substantiated by 
other data and analyses, this would indicate that the symmetry 
part of the fragment binding energy is slightly weaker than that of
isolated nuclei. Fragments, as they are formed at breakup, may have a
lower than normal density. Such effects may be enhanced as the energy
deposited in the fragmenting system is increased. Therefore,
exclusive studies with possibly heavier projectiles will be required
to more clearly identify potential variations of the symmetry energy with
the reaction parameters. 

\begin{acknowledgments}
Stimulating discussions with L. Andronenko, M. 
Andronenko, S.~DasGupta, W.G.~Lynch,
I.N.~Mishustin, M.B. Tsang and with the ALADIN group are acknowledged.
One of the authors (A.S.B.) would like to thank the GSI for warm
hospitality and support.
\end{acknowledgments}

%\newpage


\begin{thebibliography}{99}
\bibitem{notation}
to avoid confusion with the notation of \protect\cite{tsang01} the 
parameter $\beta$ used in \protect\cite{boga74,boga76,boga80} is here 
represented by $\beta_{t3}$
\end{thebibliography}
\end{document}